\documentstyle[emulateapj,psfig]{article}

\newcommand{\be}{\begin{equation}}
\newcommand{\ee}{\end{equation}}
\newcommand{\ba}{\begin{eqnarray}}
\newcommand{\ea}{\end{eqnarray}}
\newcommand{\siml}{\lower4pt \hbox{$\buildrel < \over \sim$}}
\newcommand{\simg}{\lower4pt \hbox{$\buildrel > \over \sim$}}
\newcommand{\Mesz}{M\'esz\'aros}

\slugcomment{Accepted for publication in The Astrophysical Journal,
Aug. 19, 2002}

\begin{document}

\title{An analysis of gamma-ray burst spectral break models}

\author{Bing Zhang \& Peter M\'esz\'aros}
\affil{Department of Astronomy \& Astrophysics, Pennsylvania State
    University, University Park, PA 16803}

\begin{abstract}
Typical gamma-ray burst spectra are characterized by a spectral
break, $E_p$, which for bright BATSE bursts is found to be narrowly
clustered around 300 keV. Recently identified X-ray flashes, which
may account for a significant portion of the whole GRB population,
seem to extend the $E_p$ distribution to a broader range below 40 keV.
Positive correlations among $E_p$ and some other observed parameters
have been noticed. On the other hand, within the cosmological fireball
model, the issues concerning the dominant energy ingredient of the
fireball as well as the location of the GRB emission site are still
unsettled, leading to several variants of the fireball model. 
Here we analyze these models within a unified framework, and
critically reexamine the $E_p$ predictions in the various model 
variants. Attention is focused on the predictions of the narrowness of
the $E_p$ distribution in different models, and the correlations among
$E_p$ and some other measurable observables. These model properties
may be tested against the current and upcoming GRB data, through which
the nature of the fireball as well as the mechanism and site of the
GRB emission will be identified. In view of the current data, various
models are appraised through a simple Monte-Carlo simulation, and a
tentative discussion about the possible nature of X-ray flashes is
presented.
\end{abstract}


\section{Introduction}
\label{sec:intro}

Gamma-ray burst (GRB) sub-MeV spectra contain most of the prompt
information available from these mysterious sources. Progress in
understanding the origin of such spectra, however, has been slow. A
typical GRB spectrum is non-thermal, and can be fitted by a so-called 
Band-function (Band et al. 1993) with three parameters: a low
energy power law photon index $\alpha$, a high energy power law photon
index $\beta$, and the spectral break energy $E_p$ which defines the
smooth transition between the two power laws. Since generally $\beta <
-2$, $E_p$ is related to the peak of the $\nu F_\nu$ spectrum,
and therefore is also called ``E-peak". 
Within the framework of the commonly considered cosmological fireball 
model, there are a number of variants (invoking different fireball 
contents, different emission sites, or different emission mechanisms)
proposed to explain the GRB data. An important quantity which 
characterizes the GRB is $E_p$, which for 156 bright BATSE bursts in the 
4B sample is found to be narrowly clustered around 300 keV, with a 
log-normal distribution with full width at half maximum of less than a 
decade (Preece et al. 1998; 2000). Theoretically, the value of $E_p$ is
expected to be correlated with some other observational parameters,
but such correlations could vary significantly in different
models. These provide important criteria to evaluate the correctness
of the models by comparing such correlations found in the data.

This topic has received heightened attention in view of two recent
developments. On the one hand, a new category of X-ray transients, 
known as X-ray flashes (XRFs) has been identified
(Heise et al. 2001; 2002, in preparation). The XRFs resemble normal
GRBs in many respects, with the novelty that the peak energies are
distributed from 100 keV to below 40 keV. These objects appear to form
a natural extension of the GRB population in the softer and fainter
regime (Kippen et al. 2001; 2002), which widens the narrow $E_p$
distribution found in the BATSE data, and it is estimated that XRFs
represent a large portion (e.g. $\siml 1/3$) of the whole GRB
population. It is therefore interesting to see how present theories
can accommodate this new category of GRBs. On the other hand, GRB
light-curve variabilities (Fenimore \& Ramirez-Ruiz 2000; Reichart et
al. 2001) and spectral lags (Norris et al.  2000; Norris 2002) have
been proposed as Cepheid-like luminosity indicators for the long
duration ($t_b \simg 2$ s) GRBs. These offer the exciting prospect of
a direct relation between observable quantities and some of the
theoretically most relevant parameters, such as the wind luminosity
$L$, GRB intrinsic durations, emission-site magnetic fields, as well
as the characteristic of the synchrotron or inverse Compton energies
(which in many models are linked to the observed $E_p$).
Empirically, some such correlations have been noticed. Lloyd-Ronning \&
Ramirez-Ruiz (2002) discovered a positive dependence of $E_p$ on the
GRB variability (or luminosity). When combining this with the
luminosity - variability correlation (Fenimore \& Ramirez-Ruiz 2000;
Reichart et al. 2001), one infers a positive correlation between $E_p$
and the burst luminosity $L$. Indeed, such a correlation has been seen
in the BeppoSAX bursts with known redshifts (Amati et al. 2002).
Thus, it is timely to critically revisit the physical $E_p$ predicted
in various models, as a first step towards the goal of constraining or
even identifying the nature of the fireball as well as the relevant
emission site and mechanism for the GRB prompt emission. This is the
purpose of the present paper. In \S2.1, we present a synthesis of the
current GRB model variants within a unified framework, and define the
parameter regimes in which each model variant applies. In the rest of
\S2, we revisit these model variants, focusing specifically on the
$E_p$ predictions. In \S3, these predictions, as collected in Table
\ref{tbl-2}, are used to evaluate the current models through a simple
Monte-Carlo simulation, and the possible nature of the X-ray flashes
is tentatively discussed. Conclusions are drawn in \S4.

\section{GRB models and $E_p$ predictions}
\label{sec:predic}
\subsection{GRB model synthesis}
\label{subsec:synth}

Cosmological fireball models invoke a brief release of energy $\sim
10^{51}-10^{53}~{\rm ergs}$ within a short duration of time $\sim
10^{-3}-10^3$ s. The fireball must be clean (low baryon load) so that
after initial acceleration, the fireball is relativistic.
The non-thermal spectrum requires the emission to be optically-thin,
so that the emission site should be above the photosphere defined by
the baryon content.
The fireball eventually decelerates by the interstellar medium when
the afterglow starts. The site of the GRB emission is therefore
limited to the regime
\be
r_{ph} \siml r \leq r_{dec},
\ee
where $r_{ph}$ (e.g. eq.[\ref{eq:rph+-}], see eq.[5] of \Mesz~ et
al. 2002 for a more general treatment) and $r_{dec}$
(eq.[\ref{eq:rdec}]) are the baryonic photosphere radius and the
deceleration radius, respectively.
The non-thermal nature of GRB emission also requires that the emission
energy is not directly coming from the hot fireball (which gives a
thermal-like spectrum), but derives from some other forms, e.g., the
kinetic energy of the baryon bulk or the magnetic energy of the fireball.
Generally, the GRB central engine involves a rapidly spinning and
possibly highly magnetized object such as a black hole - torus system
or a millisecond magnetar\footnote{Strongly magnetized central engines
have been invoked in GRB models motivated by their ability to launch
collimated jets and to avoid heavy baryon loading (Usov 1992; \Mesz~
\& Rees 1997b; Wheeler et al. 2000). Also, the spindown luminosity of
the rapidly rotating
central object, whether a black hole or magnetar, can be tapped through
its Poynting flux, which in many cases is more powerful than the initial
thermal luminosity of the fireball, e.g. Lee, Wijers \& Brown 2000;
van Putten 2001).}.

The fireball luminosity therefore may be broadly divided into two
components. The first component, which is the component conventionally
invoked in the simplest fireballs, is initially composed of thermal
photons, pairs and a small amount of baryons. This component, which we
call the hot component, essentially stores its energy in the form of
the kinetic energy of the baryons after initial acceleration, apart
from (usually) a small fraction of energy leaking at the baryonic
photosphere. The second component, which we call the cold component,
is carried by a Poynting-flux and low-frequency waves associated with
the central engine spin. For ease of discussion below, we broadly
define the hot and the cold luminosity components as $L_h \simeq L_K$
and $L_c \simeq L_P$, respectively, so that the total fireball
luminosity is $L=L_h+L_c=L_K+L_P$. We also define the ratio of the
cold-to-hot components as $\sigma \equiv L_c/L_h \simeq L_P / L_K$
following the convention of pulsar wind nebula theories (e.g. Rees \&
Gunn 1974;  Kennel \& Coroniti 1984)\footnote{Strictly speaking,
such a definition is equivalent to that in the pulsar wind nebula
theory only when $\sigma$ is not too large (e.g. $\sigma <
10^4$). This is because, as we know, a pure cold component (e.g. the
pulsar wind from Crab) also has a small fraction [$\sim
(10^{-5}$-$10^{-4})$] of energy stored as the kinetic energy of the pairs
flowing from the pulsar magnetosphere. In this sense, any extremely
large $\sigma$ (e.g., $\sigma_{c3}$ below, eq.[\ref{sigmac3}]) no
longer has the conventional meaning of the Poynting flux-to-kinetic
energy ratio.}. A fireball may be then defined as
Poynting-flux-dominated if $\sigma \simg 1$ or as
kinetic-energy-dominated if $\sigma\siml 1$. In principle $\sigma$
decreases with increasing radius due to conversion of part of the
Poynting flux into the kinetic energy, especially when magnetic
reconnection is operating, though the effect is likely not to be large
(Lyubarsky \& Kirk 2001; Contopoulos \& Kazanas 2002). In any case,
in the following discussions $\sigma$ refers to the specific value at
the relevant radius in the problems.

The deceleration radius is defined by the kinetic energy of the
fireball, $E_K=E/(1+\sigma)$, if the kinetic energy is decoupled from
the magnetic energy by the time of deceleration\footnote{In the MHD
regime, the kinetic energy and the magnetic energy are coupled, so the
deceleration radius is still defined by the total energy $E$ of the
fireball. Since we are interested in the comparison between $r_{dec}$
and $r_{_{\rm MHD}}$ (eq.[\ref{rbr}], at which both energy components
are decoupled), it is more relevant to define $r_{dec}$ with $E_K$.},
so that
\ba
r_{dec}=\left({3 E_K}/{4\pi n_{ext} m_p c^2 \Gamma^2}\right)^{1/3}
\nonumber \\
 =5.4\times 10^{16}~{\rm cm}~
 E_{52}^{1/3} (1+\sigma)^{-1/3} n_{ext}^{-1/3} \Gamma_2^{-2/3},
\label{eq:rdec}
\ea
where $n_{ext}$ is the density of the interstellar medium, $\Gamma$ is
the bulk Lorentz factor of the fireball before deceleration, and
hereafter the convention $Q_n=Q/10^n$ is adopted. The photosphere
radius depends on a number of factors and has been discussed in \Mesz~
et al. (2002) (see also \S2.4). For a very clean fireball in which the
opacity is defined by the thermal pairs rather than baryonic electrons,
\be
r_{ph,\pm} =r_0 (\Theta_0/\Theta'_\pm)=1.9\times
10^9~{\rm cm}~ L_{52}^{1/4} (1+\sigma)^{-1/4} t_{v,m,-3}^{1/2},
\label{eq:rph+-}
\ee
where $\Theta'_\pm \sim 0.03$ is the normalized (in unit $m_e c^2$)
comoving temperature of the fireball when the pairs drop out of
equilibrium (Shemi \& Piran 1990; Paczy\'nski 1986; Goodman 1986),
and
\be
\Theta_0 =(k/m_ec^2)(L_h /4\pi r_0^2
\bar\sigma)^{1/4}=1.9 L_{52}^{1/4} (1+\sigma)^{-1/4} t_{v,m,-3}^{-1/2}
\label{Theta0}
\ee
($\bar\sigma$ is the Stefan-Boltzmann constant, hereafter symbols like
$m_e$, $m_p$, $c$, $e$, $\hbar$, etc. denote fundamental physical
constants with conventional meanings) is the initial
temperature of the fireball at the radius $r_0\sim ct_{v,m}=3\times
10^7~{\rm cm}~t_{v,m,-3}$ ($t_{v,m}$ is the minimum variability
timescale of the central engine, which defines the width of each
mini-shell to be $c t_{v,m}$), and $L_h=L/(1+\sigma)$ is the
luminosity of the hot component.
In the case where the  baryon load is somewhat larger and
the opacity is dominated by baryonic electrons, the photosphere radius
is given instead by eq.(5) of \Mesz~ et al. (2002). In any case,
equation (\ref{eq:rph+-}) marks the radius above which the pairs are
no longer in equilibrium.

Another radius of interest is the critical radius where the MHD
treatment breaks down, i.e., $r_{_{\rm MHD}}$. A convenient way to
define $r_{_{\rm MHD}}$ is to require that the local plasma density,
$n=n_b+n_\pm$ which includes both the baryon density and the pair
density, to be equal to the Goldreich-Julian (1969, hereafter GJ)
density, $n_{GJ}$, which is the minimum density required for the
plasma to be frozen in the magnetic field (Usov 1994)\footnote{For an
alternative but intrinsically similar discussion about the MHD
condition, see Spruit, Daigne \& Drenkhahn (2001).}.  The GJ density
drops as $\propto r^{-3}$ within the light cylinder, $r_{lc}=c/\Omega$
(where $\Omega$ is the angular frequency of the central engine), but
as $\propto r^{-1}$ beyond $r_{lc}$, so that at large
distances, the GJ density is $n_{GJ}=(\Omega B_s/2\pi ec)(R/r_{lc})^3
(r_{lc}/r)= (B_s R^3 \Omega^3/2\pi e c^3 r)$, where $R$ is the radius
of the magnetized central engine. Assuming that the
spin-down energy of the central engine is mainly carried by the
Poynting flux, one has $L_P=B_s^2 R^6 \Omega^4/6c^3$. The GJ
density in the rest frame of the fireball can be then expressed in
terms of the measurable parameters
(hereafter the primed quantities denote those measured in the rest
frame of the fireball)
\be
n'_{GJ}=1.0\times 10^8~{\rm cm}^{-3}~[\sigma/(1+\sigma)]^{1/2}
L_{52}^{1/2} t_{v,m,-3}^{-1} r_{13}^{-1} \Gamma_2^{-1},
\ee
where $t_{v,m} \sim 2\pi/\Omega \sim 10^{-3}$ s, and the value has
been normalized to $r\sim 10^{13}$ cm which is the typical radius for
the ``internal'' $\gamma$-ray emission (e.g. from the internal
shocks). The comoving plasma density at
the distance $r$ is $n'=n'_b+n'_\pm$. For a continuous wind, the comoving
baryon density $n'_b=L_K/4\pi r^2 m_p c^3\Gamma^2$ is
\be
n'_b = 1.8\times 10^{13}~{\rm cm}^{-3}~
         L_{52} (1+\sigma)^{-1} r_{13}^{-2} \Gamma_2^{-2}.
\ee
The pair density is much lower than this.  While they are in
equilibrium, which is below $r_{ph,\pm}$ (eq.[\ref{eq:rph+-}]) and
generally somewhere below the baryonic photosphere, the pair
number density is $n'_\pm(th)\simeq 4.41\times 10^{30} {\rm cm}^{-3}
{\Theta'}^{3/2} \exp (-{\Theta'}^{-1})$ (Pacz\'nski 1986). At the
radius where $\Theta'_\pm \sim 0.03$, this is $n'_\pm(th)=7.6\times
10^{13}~{\rm cm}^{-3}$. Beyond the pair-freeze
radius, $n'_\pm$ drops as $\propto r^{-2}$, so that at the fiducial
radius $r_{13}=1$, one has (cf. Usov 1994)
\be
n'_\pm \sim 2.8\times 10^6~{\rm cm}^{-3}~L_{52}^{1/2}
(1+\sigma)^{-1/2} t_{v,m,-3}^{-1} r_{13}^{-2}.
\label{npm}
\ee
In principle, further annihilation of these pairs is possible. In the
context of the current problem, the annihilation time scale is much
longer than the expansion timescale, so (\ref{npm}) generally applies.
While $n'_\pm < n'_{GJ}$, the baryon density $n'_b \gg n'_{GJ}$ at
small radii, although it will drop below $n'_{GJ}$ at a large enough
radius due to the different $r$-dependence of $n'_b$ and $n'_{GJ}$.
By requiring $n'_b \sim n'_{GJ}$, one can define the radius where the
MHD approximation breaks down, i.e.,
\be
r_{_{\rm MHD}}=1.8 \times 10^{18}~{\rm cm}~L_{52}^{1/2}[\sigma(1+\sigma)]
^{-1/2} t_{v,m,-3} \Gamma_2^{-1}.
\label{rbr}
\ee
This radius is usually beyond the deceleration radius $r_{dec}$ (so
that MHD approximation never breaks) unless
\be
\sigma > \sigma_{c2}= 190 L_{52}^{3/4} E_{52}^{-1/2}
t_{v,m,-3}^{3/2} n_{ext}^{1/2} \Gamma_2^{-1/2}
\label{sigmac2}
\ee
By setting $n'_b=n'_\pm$, one can also define a critical value
\be
\sigma=\sigma_{c3}=4.1\times 10^9 L_{52}t_{v,m,-3}^2 \Gamma_3^{-4}~,
\label{sigmac3}
\ee
above which the electrons associated with baryons are negligible.

Finally, there is another critical value of
\be
\sigma=\sigma_{c1} \sim (0.1-1)~,
\label{sigmac1}
\ee
which separates the regimes where strong shocks can or cannot
develop. Strong shocks are only possible for low $\sigma$ flows. For
$\sigma > \sigma_{c1}$, the shocks are quite weak, mainly because the
three speeds on both sides of the shock are close to the speed of light
(Kennel \& Coroniti 1984).

A fireball may be classified into several sub-categories depending on
the value of $\sigma$. Table \ref{tbl-1} lists the characteristics of
different fireball variants. GRB models can be also naturally
classified into three categories based upon the proposed location of
the GRB prompt emission. Within each category, several subtypes are
defined according to the regime of $\sigma$, as follows.

1. {\em Internal models}, with $r_{ph} < r < r_{dec}$. In the standard
scenario the energy input is an unsteady, kinetic-energy-dominated
wind which undergoes dissipation in internal shocks (Rees \& \Mesz~ 1994),
which applies for $\sigma < \sigma_{c1}$. For $\sigma > \sigma_{c1}$, the
source of energy dissipation may be from the strong magnetic fields
in the fireball. For $\sigma > \sigma_{c2}$, the MHD approximation breaks
down beyond the radius $r_{_{\rm MHD}} < r_{dec}$, and a global energy
dissipation is
expected to occur. For the extreme case of $\sigma>\sigma_{c3}$, the
baryon content is negligible, and the process is analogous to a
relativistic pulsar wind (Usov 1994; Lyutikov \& Blackman 2001).
For $\sigma_{c1}<\sigma <\sigma_{c2}$, the global MHD approximation
applies all the way to the deceleration radius. Some local breakdown of
the MHD condition is required (and possible), probably through magnetic
reconnection (e.g. Drenkhahn \& Spruit 2002).

2. {\em External models}, with $r=r_{dec}$. The standard picture is
that of an external shock model (Rees \& \Mesz~1992; \Mesz~ \& Rees
1993; Dermer, Chiang \& B\"ottcher 1999). A variant of this model
invokes the magnetic wind-medium interaction in the high-$\sigma$
regime (Smolsky \& Usov 2000).

3. {\em Innermost models}, with $r \simg r_{ph}$. Although the thermal
nature of the photosphere emission seems to preclude its role as the
main GRB emission mechanism, a few bursts are known to have
quasi-thermal $\gamma$-ray spectra. A non-thermal character may also
appear in the photospheric emission through
Comptonization during the emergence of the spectrum (Goodman 1986). Also,
with a strong magnetic component (not necessarily of very high $\sigma$),
Alfv\'en turbulence induced and propagated from the deep layers beneath
the baryonic photosphere can also lead to a non-thermal Compton tail
(Thompson 1994; \Mesz~ \& Rees 2000). Even without Comptonization, both
the baryonic and the pair photosphere components can provide additional
components to the optically-thin component, and under certain circumstances
even play a dominant role, which have been suggested to be responsible for
the recently identified X-ray flashes (\Mesz~ et al. 2002).

The most natural radiation mechanism in the optically thin region
(both for the internal and the external models) is the synchrotron
radiation (or its variants, e.g., jitter radiation, synchro-Compton
radiation, random electric field radiation, etc.) and its self-inverse
Compton (IC) emission, although the acceleration mechanism and the
distribution function of the emitting electrons may vary in different
models. For the baryonic photosphere Comptonization model, the $E_p$
is essentially defined by the thermal peak although the spectral shape
is modified. Below we revisit the $E_p$ predictions in these model
variants.

\subsection{Internal models}
Generally, the characteristic synchrotron emission energy
(which is naturally connected to the $E_p$ energy in this model)
of the electrons with typical energy $\gamma_e m_e c^2$ in a
relativistic ejecta
of bulk Lorentz factor $\Gamma$ is $E_p^{syn}=(4\Gamma/3)(3/2)\gamma_e^2
(\hbar eB'/m_e c) \sin \Psi (1+z)^{-1}$, or\footnote{By adopting such a
notation, one has implicitly assumed that there is relativistic emission
beamed towards the line of sight. If the relativistic ejecta is in a jet
with a sharp cutoff at the edges, and when the viewing direction is offset
from the jet, the $\Gamma$ factor should be replaced by a more general
Doppler factor (e.g. Woods \& Loeb 1999; Nakamura 1999; Salmonson
2001). However, if the GRB jets are structured with smooth variation
of energy per solid angle in different directions (e.g. Rossi, Lazzati
\& Rees 2002; Zhang \& \Mesz~ 2002; see also Woosley, Zhang \& Heger
2002, within the context of the collapsar models), this expression has
already taken into account the viewing effect, since what matter are
simply the relevant parameters (e.g. $\Gamma$, $L$, etc.) along the
line of sight (Zhang \& \Mesz~2002).}
\be
E_p^{syn}
 \simeq 2\times 10^{-8}~{\rm eV}~(\Gamma B')\gamma_e^2 \sin \Psi
(1+z)^{-1}.
\label{Ep}
\ee
where $B'$ is the comoving magnetic fields, $\Psi$ is the mean pitch
angle of the electrons, and $z$ is the source redshift. The factor
$(\Gamma B')$ is the Lorentz-boosted magnetic
field strength, whose energy density should be proportional to the
fireball energy density in the observer's frame as long as the
magnetic energy density is a constant fraction of the total energy
density. This gives $\Gamma B' \propto U^{1/2} \propto (L/r^2)^{1/2}
\propto L^{1/2}r^{-1}$, where $U$ and $L$ are the total energy density
and the luminosity of the fireball.  In principle there could be two
components that contribute to the co-moving magnetic field in the
emission region. The wind itself usually carries some ``primordial''
magnetic field ($B'_w$) from the magnetized central engines, the
lab-frame energy density of which can be estimated as
$U_B=B_{w}^{2}/8\pi=L_P/4\pi cr^2$. The comoving field strength is
then
\ba
B'_w=[{2 L \sigma}/{(1+\sigma)c r^2}]^{1/2}/\Gamma \nonumber \\
\simeq 8.2\times 10^5~{\rm G}~[\sigma/(1+\sigma)]^{1/2} L_{52}^{1/2}
r_{13}^{-1} \Gamma_2^{-1}.
\label{Bw}
\ea
Although such a large scale wind field is
globally organized, we assume that in the dissipation regions
(internal shocks or the reconnection sites) such a field is
re-distributed randomly with comparable strength.  Alternatively,
magnetic fields could be generated in-situ ($B'_s$) if the energy
dissipation is through the internal shocks (e.g. Medvedev
\& Loeb 1999). Assuming that each proton gains a mean random energy
$\theta_p m_p c^2$ from the internal shocks, the co-moving thermal energy
density is $u'=n' \theta_p m_p c^2$, where $n'$ is the comoving baryon
number density in the shock-heated region. The parameter $\theta_p\sim
(\gamma_{ij}-1)$ depends on the relative Lorentz factor between the
shells, $\gamma_{ij} \simeq (\gamma_i/\gamma_j+\gamma_j/\gamma_i)/2$,
where $\gamma_i$ and $\gamma_j$ are the Lorentz factors of the two
colliding shells, respectively. Though uncertain, $\theta_p$ is of
order unity for typical parameters, and is not expected to be
sensitive to the absolute value of $\Gamma$. When a collision occur,
the slow shell enters the spreading phase due to the internal velocity
difference within the shell itself. In this regime, the density in the
shock-heated region evolves as $n'\simeq \zeta M_0 \Gamma/(4/3)\pi r^3
m_p \sim \zeta Lt_v/(4/3)\pi r^3 m_pc^2 \propto r^{-3}$, where $M_0=L
t_v/\Gamma c^2$ is the total mass in the colliding shell, $t_v$ is the
typical injection duration of the mini-shells, so that $\Delta_0 \sim
c t_v$ is the typical initial shell width, and $\zeta$ is the
compressive ratio (at least 7 for strong shocks). If the
shock-generated magnetic fields reach an equipartition fraction
$\epsilon_B$ of the total thermal energy in the shock heated region,
the co-moving field strength in the forward shocked region (in the
slow shell) is then
\ba
B'_s =(8 \pi \epsilon_B n' \theta_p m_p c^2)^{1/2} \nonumber \\
\simeq 1.4 \times 10^6 ~{\rm G}~(\zeta_1
\epsilon_{B,-1}\theta_p)^{1/2} L_{52}^{1/2} r_{13}^{-1}
\Gamma_2^{-1},
\label{Bs}
\ea
where $r\sim \Gamma^2
\Delta_0$ (the internal shock radius) has been used.
By comparing $B'_w$ and $B'_s$, we find that the wind component and
the shock component of the magnetic fields have a similar amplitude
for reasonable parameters, and share the same dependences on the
parameters such as $L_{52}$, $r_{13}$ and $\Gamma$. This is
understandable, since both the wind magnetic density and the shock
magnetic density are assumed to be a constant fraction of the total
energy density of the fireball (through $\sigma$ and $\epsilon_B$,
respectively). For typical parameters, the co-moving field strength is
$B'=B'_w+B'_s = 10^6 ~{\rm G}~\epsilon_{x1} L_{52}^{1/2} r_{13}^{-1}
\Gamma_2^{-1}$, where $\epsilon_{x1}=0.82[\sigma/(1+\sigma)]^{1/2}+
1.4(\zeta_1\epsilon_{B,-1} \theta_p) ^{1/2}$ for the internal shock
models, or $\epsilon_{x1}=0.82[\sigma/(1+\sigma)]^{1/2}$ for the
magnetic field dissipation models.  We then have
\be
\Gamma B'=10^8 ~{\rm G}~\epsilon_{x1} L_{52}^{1/2} r_{13}^{-1},
\label{B}
\ee
which removes the apparent $\Gamma$-dependence (although $r_{13}$ is
still dependent on $\Gamma$ in the internal shock models). Notice that
even if $\epsilon_B \ll 1$ (as shown in the afterglow fits),
$\epsilon_{x1}$ can be still close to unity given a large enough wind
component (e.g. $\sigma > 0.1$). Equation
(\ref{B}) is essentially model-independent.

The characteristic Lorentz factor of the emitting electrons $\gamma_e$,
however, depends on the poorly understood mechanisms of particle
acceleration. Conventionally, given a typical magnetic field at the
acceleration site (eq.[\ref{B}]), there are two characteristic
$\gamma_e$'s from simple scalings. One is defined through the simple
equipartition argument, i.e., to parameterize the random electron
energy density to be proportional to the comoving thermal energy. Due
to usually the same scaling laws of both densities, the typical
electron Lorentz factor obtained using this method, denoted as
$\gamma_e^{(1)}$, generally does not depend on the radius of emission
and particle acceleration. A second estimate of $\gamma_e$, denoted as
$\gamma_e^{(2)}$, is
obtained by limiting the maximum gyro-acceleration at the local
magnetic field by the synchrotron (and/or IC) cooling. For the
simplest case, the typical acceleration time is $t'_a=2\pi \gamma_e
(m_e c/e B')$, and the typical cooling time is $t'_c=(\gamma_e m_e
c^2)/[(4/3)\gamma_e^2 \sigma_{T}c({B'}^2/8\pi)]$, where $\sigma_T$ is
the Thomson cross section. Solving $t'_a=t'_c$, the maximum achievable
electron Lorentz factor is
\be
\gamma_{e}^{(2)}=(3/2)m_e c^2/[(2\pi)^{1/2} e^{3/2}
B'^{1/2}] = 4.7\times 10^4 {B'}_6^{-1/2}.
\label{game2}
\ee
This is radius-dependent (through different $B'$ at different radii),
but in such a form (e.g. $\gamma_e^{(2)}\propto {B'}^{-1/2}$) that the
comoving typical synchrotron frequency only depends on fundamental
parameters. In the real problems, which of the two typical $\gamma_e$'s
($\gamma_e^{(1)}$ or $\gamma_e^{(2)}$) is relevant depends on the
concrete conditions involved, and on the particle spectrum. Below we
will discuss both the low $\sigma$ and high $\sigma$ cases for the
internal models.

\subsubsection{Low $\sigma$: internal shock dissipation}
For the internal shock model, electrons are accelerated at the shocks
via the Fermi mechanism to achieve a power-law distribution with the
index $-p$. The minimum Lorentz factor of the injected electrons
(i.e. $\gamma_e^{(1)}$ in the above notation) is
$\gamma_e=f(p)({m_p}/{m_e}) ({\epsilon_e}/{\xi_e})\theta_p
\simeq  310\epsilon_{x2}$,
where $\epsilon_e$ is the electron equipartition factor which is
believed to be close to unity (e.g. 0.3) in the internal shocks,
$\xi_e$ is the injection fraction of the electrons which could in
principle be less than unity (Bykov \& \Mesz~ 1996; Daigne \&
Mochkovitch 1998), $f(p)=(p-2)/(p-1)$, and the typical value $1/6$ as
$p=2.2$ has been adopted to get the numerical value. The parameter
$\epsilon_{x2} =[{f(p)}/{(1/6)}] ({\epsilon_e}/{\xi_e}) \theta_p$
could be of order unity, if $\xi_e$ is below unity for more than
an order of magnitude. One important feature of the internal shock
model is that $\theta_p$, and therefore $\gamma_e$ only depends on the
relative Lorentz factor between the colliding shells, which should not
strongly depend on the actual bulk Lorentz factor of the shell,
$\Gamma$, or the radius of the emission region, $r\propto \Gamma^2$.
Notice that in this paradigm, $\gamma_e^{(2)}$ (eq.[\ref{game2}])
defines the maximum acceleration energy of the electrons, and is much
larger than $\gamma_e^{(1)}$ ($\sim 310$), and $\gamma_e^{(1)}$ is
relevant in the problem. Re-writing (\ref{Ep}), we get the peak energy
of the internal shock synchrotron model,
\be
E_p^i(low~ \sigma)
\simeq  200~{\rm keV}~\epsilon_{x3} (L_{52}^{1/2}
r_{13}^{-1}) (1+z)^{-1},
\label{Ep1}
\ee
where $\epsilon_{x3}=\epsilon_{x1} \epsilon_{x2}^2 \sin \Psi$.
It is worth emphasizing that the $r^{-1}$ dependence in (\ref{Ep1})
infers that $E_p \propto \Gamma^{-2}$ (see also Ramirez-Ruiz \&
Lloyd-Ronning 2002).

In some cases, e.g., long $t_v$ (large $r$) or large $\xi_e$ (so that
$\epsilon_{x2}$ is much below unity), the synchrotron peak is below the
BATSE window, and the synchrotron self-inverse Compton component may
be responsible for the BATSE emission (Panaitescu \& \Mesz~ 2000). The
peak energy in this model is $E_p^{IC}=\gamma_e^2 E_p^{syn}$, or
\be
E_p^{i,IC}(low~\sigma)
\simeq  200~{\rm keV}~\epsilon_{x4} (L_{52}^{1/2}
r_{13}^{-1}) (1+z)^{-1},
\label{Ep2}
\ee
where $\epsilon_{x4}=\epsilon_{x1} (\epsilon_{x2}/0.06)^4 \sin \Psi$. In both
eqs.(\ref{Ep1}) and (\ref{Ep2}), all the parameters related to the
shock physics are absorbed into two single parameters, $\epsilon_{x3}$,
$\epsilon_{x4}$, and the dependence, $E_p \propto L_{52}^{1/2}
r_{13}^{-1} (1+z)^{-1}$ is valid for both the the
synchrotron and IC components. Several caveats ought to be
made for the IC-dominated spectrum. First, $\epsilon_B/\epsilon_e \ll
1$ is generally required to achieve a prominent IC component (Zhang \&
\Mesz~ 2001 and references therein). In the internal shock regions,
the effective $\epsilon_B$ can not be too low below unity given a not
too small wind component $B'_w$ (e.g. $\sigma > 0.1$). Second, an
IC-dominant fireball is inefficient in converting energy to radiation
since multi-IC components at even higher energies consume much more
energy so that one requires a very large energy reservoir to begin with
(Derishev et al. 2001). Third, since $\epsilon_{x3} \propto
\epsilon_{x2}^2$ while $\epsilon_{x4} \propto \epsilon_{x2}^4$, the IC
mechanism tends to produce a wider $E_p$ distribution than the
synchrotron mechanism (see \S3 and Fig.1 for more discussions).

\subsubsection{High $\sigma$: magnetic dissipation}
If a wind has a high $\sigma > \sigma_{c1}\sim (0.1-1)$, strong
internal shocks may not occur. Alternatively, intense energy
dissipation may occur through magnetic reconnection and/or plasma
instability. For $ \sigma_{c1}< \sigma < \sigma_{c2} \sim 190$, MHD
does not break globally, but local dissipation is possible through
reconnection. If reconnection occurs below the photosphere, the
dissipated energy is used to accelerate the fireball bulk. Only when
$\sigma$ is above some value could the energy dissipation occur beyond
the photosphere (Drenkhahn 2002). For $\sigma > \sigma_{c2}$, the MHD
approximation breaks down globally before the deceleration radius, and
a more violent, global energy dissipation is possible. The particle
acceleration mechanism in both cases is hard to delineate.
For the local dissipation case, the in-principle-achievable electron
energy may be defined by the DC component of the electric field, i.e.,
$\gamma_{e,M,DC} m_ec^2=e E' l'$, where $l'\sim \Gamma c/\Omega \sim
10^9 \Gamma_2 l_7$ is typical comoving scale of the stripped wind
which may be regarded as the characteristic reconnection scale, $E' =
\alpha (v_{A}/c) B'$ is the comoving DC electric field, where $v_A\sim
c$ is the Alfv\'en speed, and $\alpha$ is a poorly constrained
parameter (Syrovatskii, 1981; Craig \& Litvinenko 2002). This gives
$\gamma_{e,M,DC} \sim 6\times 10^{11} \alpha$ for typical values,
which is extremely high unless $\alpha$ is very small. In reality,
such a linear acceleration would be disrupted by any perturbation
perpendicular to the DC direction, and the maximum achievable
$\gamma_e$ may be still determined by the unavoidable perpendicular
magnetic fields as well as the radiation self-reaction. Estimating the
limiting $\gamma_{e}$ is difficult due to many uncertainties
involved. In any case, if the electric and magnetic fields in the
dissipation region could be regarded as quasi-random, the situation
would be quite similar to the case of the global magnetic field
dissipation, which we discuss below.

For $\sigma > \sigma_{c2}$, MHD breaks down at $r_{_{\rm MHD}} < r_{dec}$,
and magnetic dissipation occurs globally. One possible consequence is
that the MHD wind is converted to a large amplitude electromagnetic
wave (LAEMW). The electrons may then surf and get accelerated in such
a LAEMW (Gunn \& Ostriker 1971; Michel 1984; Usov 1994; Chen, Tajima
\& Takahashi 2002).
Alternatively and more plausibly, the LAEMW will soon evolves non-linearly
and induces an electromagnetic turbulence through an overturn instability,
so that the electromagnetic fields in the dissipation site are
randomized (Lyutikov \& Blackman 2001). Particles are stochastically
accelerated and emit in the random fields. To estimate the maximum
electron energy gained in the random field, we can use the two
constraints that lead to the estimates $\gamma_e^{(1)}$ and $\gamma_e^{(2)}$,
respectively. The first constraint is energy conservation. By assuming
that a fraction $\epsilon_{e'}$ of the co-moving local magnetic field
energy density is eventually converted to the co-moving kinetic energy
of the electrons, one can always write
\be
\gamma_e m_e c^2 \xi_e (n'_b+n'_\pm)=\epsilon_{e'} {B'}^2/8\pi,
\label{energy}
\ee
where $n'_b$ and $n'_\pm$ are the comoving baryon (and hence, baryon
associated electron) and pair number
densities, respectively. As discussed in \S2.1, for the typical
$\sigma$ of interest, $n'_b \gg n'_\pm$. Without generating additional
pairs (which seem to be likely as discussed below), the emitting
electrons are only those associated with the baryons, and this finally
gives $\gamma_e^{(1)}\sim 2.8\times 10^5 \epsilon_{x1}^2 \epsilon_{e'}
\sigma_2 \xi_e^{-1}$, which has no apparent $r$- or
$\Gamma$-dependences since $n'\propto {B'}^2 \propto
(L'/r^2)$. Comparing to the typical $\sim 310$ in the internal shock
models, this $\gamma_e$ is too large to reproduce the observed $E_p$
in the synchrotron model, unless $\epsilon_{e'} \sim 10^{-3} \ll 1$.

The second constraint due to radiation self-reaction would be more
relevant in this case, since the electrons are directly
accelerated in the random electric magnetic fields.
A similar estimate as (\ref{game2}) can be made. In a random
electromagnetic field, the acceleration due to the electric field
parallel to the moving direction ($E_\parallel$) is much smaller than
that due to the electric field perpendicular to the moving direction
($E_\perp$) so that the whole field may be
regarded as a random magnetic field with effective strength $\hat
B'\sim \sqrt{2} B'$ (Landau \& Lifshitz 1975; Lyutikov \& Blackman
2001). Substituting $B'$ by $\hat B'$ in (\ref{game2}), the maximum
achievable electron Lorentz factor is $\gamma_{e}^{(2)} \sim 3.9\times
10^4 {B'}_6^{-1/2}$.  This gives the typical comoving synchrotron
frequency ${E'}_{syn}\sim 30 ~{\rm MeV}$. Obviously, this is too high
to meet the observations.

There are two possibilities. First, it could be that the turbulence
scale is much less than the radiative scale $ct'_c/\Gamma$,
so that the electrons never reach the maximum acceleration, but are
limited by the turbulence scale itself. This effectively lowers the
achievable $\gamma_e$ by a constant factor without modifying the
dependence laws (Lyutikov \& Blackman 2001). One may assume
\be
\gamma_e =\epsilon_{c} \gamma_{e}^{(2)}=3.9\times 10^2
\epsilon_{c,-2}{B'}_6^{-1/2},
\label{game}
\ee
so that
\be
E_p^i(high~\sigma) \simeq 300 ~{\rm keV} \epsilon_{x5} \Gamma_2
(1+z)^{-1}.
\label{Ep3}
\ee
where $\epsilon_{x5}= \epsilon_{c,-2} \sin \Psi$. The drawback of this
possibility is that the process is still inefficient, since a low
$\epsilon_{c}\sim 10^{-2}$ again implies a very low $\epsilon_{e'}\sim
1\%$ given $n'_\pm \ll n'_b$. This is in contrast to the motivation of
the magnetic dissipation model (e.g. Drenkhahn \& Spruit 2002) which aims
to overcome the low radiation efficiency encountered in the internal
shock models. The main reason for the low efficiency in this case is
that for a higher $\sigma$ wind, there are less baryon-associated
electrons emitting given a same total wind luminosity.

The second (more plausible) possibility is that the primary electrons
do get accelerated close to $\gamma_e^{(2)}$ and emit
the $\sim 30$ MeV photons in the co-moving frame. These photons will
interact with the low energy photons (comoving energy $\sim 10$ keV)
to produce pairs if the radiation density is high enough. The new
pairs are accelerated via the same mechanism, and a pair cascade
develops, until eventually the pair density $n'_\pm =\xi'_\pm n'_b \gg
n'_b$, which limits the achievable $\gamma_e$ to be below
$\gamma_{e}^{(2)}$ through the energy budget constraint
(\ref{energy}). Since there are more leptons emitting (most being
pairs), a higher radiation efficiency is achievable. Though the
cascade process is hard to describe analytically, one may estimate
the minimum multiplicity by demanding the typical comoving emission
energy to be below the pair threshold, i.e., $\gamma_e^2 (\hbar
eB'/m_e c) \siml 0.5$ MeV. This gives $\gamma_e = (0.5/30)^{1/2}
\gamma_{e}^{(2)} \sim 5.0\times 10^3{B'}_6^{-1/2}$. Assuming a high radiation
efficiency, i.e., $\epsilon_{e'} \sim (1/3)$ in eq.(\ref{energy}), the
required minimum pair multiplicity is $\xi'_\pm=19 \sigma_2 \epsilon_{e',1/3}
L_{52}^{1/4} r_{13}^{-1/2} \Gamma_2^{-1/2}$. This still results in too
high an $E_p$. In reality, more pairs would be further produced as long
as there is a substantial hard energy spectrum extended above the
typical synchrotron energy, which would degrade the expected $E_p$
further. Notice that as long as optically thin to the pairs,
the dependence $\gamma_e  \propto {B'}^{-1/2}$ is still valid, mainly
because both the comoving pair threshold itself ($\sim 0.511$ MeV) and
the comoving characteristic synchrotron frequency ($\sim 30$ MeV for
the first generation) are constant.
The expression (\ref{Ep3}) is still valid, with $\epsilon_{x5}$ being
defined differently. The attractive feature of such a model is that
the final $E_p$ scatter is only due to the dispersion of the bulk
Lorentz factor $\Gamma$ and a parameter $\epsilon_{x5}$. This is
favorable for the observed narrow $E_p$ distribution for the bright
BATSE bursts (see \S3 and Fig.1 for more discussions).

In the magnetic dissipation model, there could be also a self-IC
component. However, since the synchrotron component already has too
high an energy for the GRB emission, invoking IC as the GRB mechanism
has to introduce some ad hoc assumptions. Also all the
criticisms to the IC internal shock models apply here as well. We
therefore do not discuss such a possibility further.

\subsubsection{Optically-thick pair photosphere}

For both the shock dissipation and the magnetic dissipation scenarios
discussed above, there could be the possibility that secondary pairs are
so abundantly generated that they form an optical thick screen for
the emergence of the photons (e.g. Guetta, Spada \& Waxman 2001;
Kobayashi, Ryde \& MacFadyen 2002; \Mesz~et al. 2002). Therefore the
observed GRB emission, at least for some cases, has undergone Compton
multi-scattering. The emergent GRB spectrum is then expected to be
regulated by the existence of the pairs. The existence of the pairs
may smear out or even destroy the ``clean'' correlations discussed
above. Due to the non-linear nature of the problem, simple analytic
$E_p$ dependences (i.e., the substitution of eqs.[\ref{Ep1}] and
[\ref{Ep3}]) are hard to provide without numerical simulations. In any
case, \Mesz~ et al. (2002) have shown that pairs tend to be
self-regulated at a moderate optical depth of $\tau'_\pm \sim$ a few,
and the co-moving typical frequency $E'_p \propto T'_\pm/{\tau'}_\pm^2$.
Since the comoving effective pair temperature is insensitive to the
shock radius as long as copious pairs are produced, such a case is
analogous to the high-$\sigma$ magnetic dissipation case, but due to
quite different reasons. The dependence $E_p \propto \Gamma
(1+z)^{-1}$ is more or less retained, with the possible weak
dependence on $r$ through $T'_\pm$. More detailed numerical work is
underway to test such a simple treatment (Ramirez-Ruiz et al. 2002, in
preparation).

The condition for forming such a pair photosphere is subject
to further investigation. Pilla \& Loeb (1998) numerically simulated
the pair production process in the internal shocks with certain model
parameters, and found that the pair processes mainly distort the high
energy part of the spectrum, with the low-energy synchrotron peak
almost unaltered. This hints that at least for some parameter regimes
(which may be large), the optically-thin synchrotron model (both
low-$\sigma$ and high-$\sigma$ cases as discussed in \S2.2.1, and
\S2.2.2) applies.
Another caveat is that a pair-regulated emission component tends to
have smoother lightcurves, since short time scale variability will be
smeared out through multi-Compton-scattering (see, e.g. Lazzati 2002).
This hints that at least for those skipy lightcurves with short time
variability, such an emission component is not important.

\subsection{External models}
The physical conditions at the deceleration radius, in certain
circumstances, also allows emission with typical energy in the BATSE
band, both for the low $\sigma$ and the high $\sigma$ cases. The main
difference for both cases is the origin of the magnetic fields in the
energy dissipation regions.

For a low-$\sigma$ flow (e.g. $\sigma < 0.1$), the picture is the
familiar external shock model (\Mesz~\& Rees 1993). The comoving
magnetic field is generated in-situ through turbulent motions to
some fraction of the equipartition value, $B'=(32 \pi \epsilon_B n_{ext}
m_p c^2 \Gamma^2)^{1/2} \simeq 40~{\rm G}~\epsilon_B^{1/2} n_{ext}^{1/2}
\Gamma_2$. The typical electron energy is $\gamma_e=[(p-2)/(p-1)]
(m_p/m_e) (\epsilon_e/ \xi_e) \Gamma=3.1\times 10^4 (\epsilon_e/\xi_e)
[f(p)/(1/6)] \Gamma_2$, where $\Gamma$ is the Lorentz factor of the
forward shocked region. The $E_p$ prediction in this model is
\be
E_p^e(low~\sigma)=75~{\rm keV}~ \epsilon_{x6} n_{ext}^{1/2}
\Gamma_2^4(1+z)^{-1},
\label{Ep4}
\ee
where $\epsilon_{x6}=\epsilon_B^{1/2}(\epsilon_e/\xi_e)^2
[f(p)/(1/6)]^2 \sin \Psi$.
It is interesting to note that $\Gamma$ appears to the fourth-power,
which can magnify any small dispersion in $\Gamma$ and tend to broaden
the $E_p$ distribution (see \S3 and Fig.1 for more discussions).
Another emission component in the low-$\sigma$ external shock scenario
is emission from the reverse shock propagating into the ejecta.
Compared with forward shocked region, the reverse shocked region has
the same comoving magnetic field strength and the bulk Lorentz factor,
but the typical random Lorentz factor is $\gamma_e^r=[(p-2)/(p-1)]
(m_p/m_e) (\epsilon_e/ \xi_e) \Gamma^r=3.1\times 10^2 (\epsilon_e/\xi_e)
[f(p)/(1/6)] \Gamma^r$, where $\Gamma^r \simeq \Gamma_0/2\Gamma$ is the
Lorentz factor of the reverse shock in the rest frame of the
undecelerated ejecta with the bulk Lorentz of $\Gamma_0$ (see
e.g. \Mesz~ \& Rees 1997a; Sari \& Piran 1999). The typical synchrotron
frequency is therefore degraded with respect to that in the forward
shock by a factor of $\Gamma^2$, and lies in the optical
band for typical parameters. However, the IC emission off these soft
photon fields by the electrons both in the reverse shock and the
forward shock regions can result in MeV photons which may (partially)
account for the GRB prompt emission (see Wang, Dai \& Lu 2001 for
detailed discussions). In any case, these IC components have high
powers of the random Lorentz factors (either $\propto
\Gamma^2(\Gamma^r)^4$ or $\propto \Gamma^4(\Gamma^r)^2)$, and will
have a similar or even broader $E_p$ distribution as the standard
external shock model (eq.[\ref{Ep4}]). Hereafter we do not explicitly
discuss these possibilities.

For the high-$\sigma$ case, the external shock variant is given by
the interaction between a relativistic plasma stream and a magnetic
barrier. As simulated by Smolsky \& Usov (2000), the outcome is
dependent on the energy density ratio between the plasma and the
field, $\alpha=8\pi n_{ext} m_p c^2 (\Gamma-1)/B'^2$. In such a case,
the comoving magnetic field $B'$ is dominated by the primordial wind
component at the deceleration radius, i.e., $B'\sim 180~{\rm G}~
L_{P,52}^{1/2} E_{K,52}^{-1/3} \Gamma_2^{-1/3} n_{ext}^{1/3}$. The
energy of electrons accelerated in the electric field generated at the
front of the
magnetic barrier is typically $\gamma_e \sim (m_p/m_e) \Gamma =
1.8\times 10^5 \Gamma_2$. The typical synchrotron energy is too high,
and Smolsky \& Usov (2000) invoke the synchro-Compton radiation in a
LAEMW as the GRB radiation mechanism. The latter is essentially a
synchrotron mechanism except that the comoving $B'$ is now replaced as
the amplitude of the LAEMW, which is $B'_w \simeq 0.1 \epsilon_{w,-1}
B'$, and that the typical electron energy is $\gamma_e \sim 200
\epsilon_{\gamma}\Gamma \sim 2\times 10^4 \epsilon_{\gamma}
\Gamma_2$. This gives the right $E_p$ energy
\be
E_p^e(high~\sigma) \simeq 880~{\rm keV}~ \epsilon_{x7}\Gamma_2^{8/3}
L_{52}^{1/2} E_{52}^{-1/3} n_{ext}^{1/3} (1+z)^{-1},
\label{Ep5}
\ee
where $\epsilon_{x7}=\epsilon_{w,-1}\epsilon_{\gamma}\sigma^{1/2}
(1+\sigma)^{-1/6}$.
The main difference between (\ref{Ep5}) and (\ref{Ep4}) is the origin
of $B'$. The wind $B'$ is determined by the properties of the central
engine, while the shock generated $B'$ depends on $\Gamma$. Thus the
power of $\Gamma$ in (\ref{Ep5}) is 4/3 lower than that in
(\ref{Ep4}), leading to less dispersion in $E_p$ for the same $\Gamma$
scatter (see Fig.1).

In both cases, pair production is likely not to be crucial in regulating
the $E_p$, due to a much smaller compactness parameter involved.

\subsection{Innermost models}
Regardless of the $\sigma$ value, a prompt signal is emitted as
soon as the fireball becomes Thomson thin. The typical energy of such
a baryonic photosphere emission is defined by the photosphere temperature,
which depends on the dimensionless entropy of the fireball. Our treatment
below follows \Mesz~ et al. (2002), with the modification that we include
a Poynting flux component. Since this cold component is not in the form
of photons initially, it is left out in calculating the temperature
of the baryonic photosphere.

The compactness parameter of the fireball can be expressed as a
dimensionless parameter\footnote{Conventionally, $m_e$ rather than
$m_p$ is used to define $\ell$, e.g. Guibert, Fabian \& Rees (1983).},
$\ell_{p,o} =(L\sigma_T/4\pi m_pc^3 r_0) \simeq 1.2\times 10^{12}
L_{52} r_{0,7}^{-1}$. For the purpose of the following discussion, $L$
should be substituted by the ``hot'' component, $L/(1+\sigma)$. The
photosphere radius is defined by the Thomson-thin condition, i.e.,
$\sigma_T n'l'=1$, where $l'$ is the comoving length of the continuous
wind or the discrete shell. There are two critical values of the
dimensionless entropy in the problem, both of which are related to
$\ell_{p,o}$. The first value $\eta_{c1} =\ell_{p,o}^{1/5}=2.5\times
10^2 [L_{52} (1+\sigma)^{-1} r_{0,7}^{-1}]^{1/5}$ is the critical
entropy below which the opacity is defined by a continuous wind rather
than a discrete shell. The second value $\eta_{c2}= \ell_{p,o}^{1/3}=
1.0\times 10^4 [L_{52} (1+\sigma)^{-1} r_{0,7}^{-1}]^{1/3}$ is the
critical entropy above which the photosphere occurs in the
acceleration regime. The observer-frame photosphere temperature
satisfies $\Theta_{ph}/\Theta_0=(\eta^{8/3} \eta_{c1}^{-10/3},
\eta/\eta_{c2},1)$ for $(\eta < \eta_{c1}, \eta_{c1}<\eta<\eta_{c2},
\eta>\eta_{c2})$, respectively (eq.[6] in \Mesz~ et al. 2002, see
detailed derivation given there). If the GRB emission is due to
Comptonization of the photospheric emission by the Alfv\'en turbulence
(Thompson 1994), the resultant $E_p$ should resemble the thermal peak
predicted by $\Theta_{ph}$. Noticing (\ref{Theta0}), we get
\ba
E_p(ph) (1+z)= 2.8 \Theta_{ph} m_e c^2 = \nonumber \\
\cases{11~{\rm keV}~L_{52}^{-5/12}(1+\sigma)^{5/12}
t_{v,m,-3}^{1/6}\Gamma_2^{8/3}, ~&   ~$\eta <\eta_{c1}$; \cr
370~{\rm keV}~L_{52}^{-1/12}(1+\sigma)^{1/12}t_{v,m,-3}^{-1/6}
\Gamma_3, ~&   ~$\eta_{c1}<\eta<\eta_{c2}$; \cr
2.7~{\rm MeV}~L_{52}^{1/4} (1+\sigma)^{-1/4} t_{v,m,-3}^{-1/2},
~& ~$\eta>\eta_{c2}$.
}
\label{Ep6}
\ea
This is an extension of the discussion in Thompson (1994), who
only discussed the highest $\eta$ case. The total photosphere
luminosity scales with the temperature, i.e. $L_{ph}/L_0=\Theta_{ph}
/\Theta_0$. Its relative importance with respect to the internal
component depends on the regimes where $\eta$ lies. For the
low-$\sigma$ case, the internal shock component dominates when $\eta$
is small, e.g. $\eta < \eta_{c1}$, but the photosphere component
starts to be prominent when $\eta > \ell^{1/4} \sim
10^3(1+\sigma)^{-1/3}$.

\section{Testing models against data}
\label{sec:testing}

We consider the following $E_p$-related observational properties: 

(i) For the 5500 spectra of 156 bright BATSE bursts analyzed
in Preece et al. (2000), the typical $E_p$ is 300 keV, and the
distribution is a narrow lognormal with full-width at half-maximum
of less than a decade. Since each burst has many spectra in this dataset,
the result implies that the $E_p$ dispersion for a particular burst
is also very narrow.

(ii) A substantial population of X-ray flashes has been recently studied,
with typical $E_p \siml $ 40 keV (Heise et al. 2001; Kippen et al. 2001).
If these events have the same origin as normal GRBs, they
extend the $E_p$ distribution to a wider range, but the small sample
collected so far indicates that the log-normal shape seem to be violated
(Kippen et al. 2002).

(iii) Statistically, a positive correlation between $E_p$ and the
isotropic luminosity $L$, $E_p \propto L^\delta$ with $\delta \sim
0.5$, has been noted recently (Amati et al. for the BeppoSAX bursts,
and Lloyd-Ronning \& Ramirez-Ruiz 2002 for the BATSE bursts).

These properties, along with other information such as temporal
variability, spectral lags, etc. are relevant quantities to be accounted
for in a theoretical model. Below we will test the model $E_p$
predictions with the above observational facts. The model predictions
are collected in Table \ref{tbl-2}.

\subsection{Narrowness of $E_p$ distribution}

To compare the narrowness of the $E_p$
distributions among different models, we performed a simple Monte
Carlo simulation. Since we are concerned about the narrowness of the
distribution rather than the absolute value of $E_p$ in different
models (and the uncertain free parameters make it possible for
all the models to adjust the absolute $E_p$ values to match the 300
keV typical value), we have adopted arbitrary units for the
parameters so that $\log E_p$ peaks around 0. We randomly
generated 40000 bursts with distributions of the unknown parameters as
follows: (1) burst luminosities span three orders of magnitude
with a $N(L)dL \propto L^{-2}dL$ distribution (such a luminosity
function is the natural consequence of the universal jet model,
Zhang \& \Mesz~2002, Rossi et al. 2002; and has received preliminary
support from the data, Schaefer, Deng \& Band 2001; Schmidt 2001);
(2) burst Lorentz factors span 1.5 orders of magnitude and are
correlated with the luminosity as $\Gamma \propto L^{1/2}$ with random
deviations (a positive $\Gamma-L$ correlation is expected in both the
conventional and the universal jet models, e.g. Kobayashi et al.
2002; Rossi et al. 2002; Zhang \& \Mesz~ 2002, and here we adopt the
(1/2) power as an example); (3) burst redshifts are randomly distributed 
in the $(0-15)$ range (a more detailed distribution function following 
the star formation history of the universe essentially does not 
influence the width of the final $E_p$ distribution, introducing only
a minor distortion of the distribution shape); (4) $\log t_v$, $\log
(T'_\pm/{\tau'_\pm}^2)$, $\log \epsilon_{x5}$,  
$\log \epsilon_{x6}$, $\log \epsilon_{x7}$, are randomly distributed 
between (0,1); (5) $\log n_{ext}$ is randomly distributed between (0,2), 
and the total duration $\log T$ is randomly distributed between (0,3); 
(6) $\log \epsilon_{x1}$ and $\log \epsilon_{x2}$ follow a standard 
Gaussian distribution with random deviations. (This is because the 
product of at least three random variables has a log-normal distribution, 
Ioka \& Nakamura 2002). For comparison, we also let 
$\epsilon_{x3}=\epsilon_{x1} \epsilon_{x2}^2$ be represented by a unit 
lognormal itself for another realization. The $E_p$ predictions within 
the different models are calculated according to Table \ref{tbl-2}. 
The fluxes of the bursts are calculated according to the standard 
$(\Omega_m, \Omega_\Lambda)=(0.3, 0.7)$ cosmology. Given a certain flux 
threshold, the histograms of the $E_p$ distributions for different 
models are plotted for comparison.

\begin{figure}
\centerline{\psfig{file=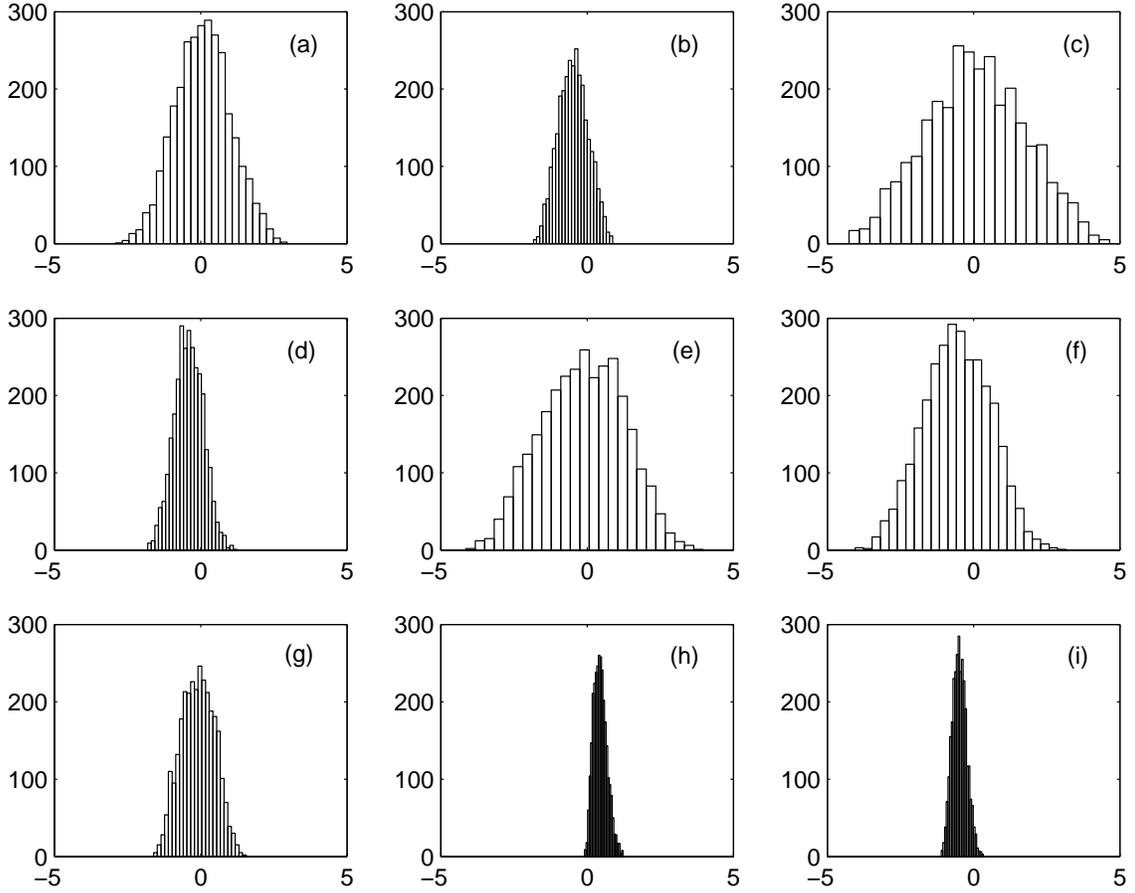,height=12.0cm}}
\caption{$E_p$ distributions from  Monte Carlo simulations for different 
models: (a) internal shock synchrotron model; (b) refined internal shock
synchrotron model with a lognormal distribution of $\epsilon_{x3}$;
(c) internal shock IC model; (d) internal magnetic dissipation model,
or pair-dominated model; (e) external shock model; (f) external
magnetic model; (g) baryonic photosphere model in the wind coasting
regime; (h) baryonic photosphere model in the shell coasting regime;
(i) baryonic photosphere model in the acceleration regime.
The input parameters are described in the text. 
The histograms represent 2962 bursts whose fluxes are within the top 
three orders of magnitude, selected from a total 40000 bursts. 
Normalized $E_p$ values are used, with typical $\log E_p \sim 0$. 
}

\label{fig1}
\end{figure}

Figure \ref{fig1} shows the $E_p$ distributions of various models.
To take into account the flux threshold effect, only the bursts whose
fluxes are among the top-three flux decades are selected (2962 bursts
from a total of 40000 bursts simulated).  For ease of comparison, we 
have plotted all the histograms with the same scale (which spans 10 
orders of magnitude). The models in the different subplots are: 
(a) the internal shock synchrotron model, with both $\epsilon_{x1}$ 
and $\epsilon_{x2}$ distributed as a lognormal; (b) the same internal 
shock synchrotron model, but with $\epsilon_{x3}$ itself assumed to be 
a standard lognormal. (This would imply that the dispersions among
$\theta_p$, $\epsilon_e$, $\epsilon_B$, etc. are small due to some
unknown conspiracy); (c) the internal shock synchrotron-self-IC model
corresponding to the synchrotron model (a); (d) the internal magnetic
dissipation model, which also applies to the pair-dominated internal 
models; (e) external shock model; (f) external magnetic model; 
(g) baryonic photosphere model in the coasting wind regime; 
(i) baryonic photosphere model in the coasting shell regime; 
(j) baryonic photosphere model in the acceleration regime.

From Fig.\ref{fig1} the following conclusions can be drawn. 
1. Given the assumed parameter distributions, neither the internal nor
the external models can reproduce the narrow $E_p$ distribution found 
in the sample of bright BATSE bursts (Preece et al. 2000). In our 
simulations, it is found that the narrowness of the $E_p$ distribution 
depends on the number of the independent parameters involved in the model, 
the narrowness of the assumed distribution of each parameter, as well as 
the power of these parameters. The more independent parameters, or the 
higher the power to which a certain parameter is involved in a model, 
the broader is the resulting $E_p$ distribution. 
2. The IC models (Fig.1c) and the external models (Fig.1e and 1f) are 
even less favored, due to the much wider distributions caused by high 
powers of some parameters in their $E_p$ expressions (e.g. $\propto 
\epsilon_{x2}^4$ for the internal shock IC model, and $\propto \Gamma^4$ 
in the external shock model). 
3. The usual internal shock synchrotron model (Fig.1a) also involves 
many parameters, and therefore results in a broad $E_p$ distribution 
among bursts. However, assuming that some of the parameters are not 
independent, one can reduce the number of free parameters to obtain 
a narrower distribution (Fig.1b). 
4. Both the internal magnetic model and the pair-dominated model invoke 
the least independent parameters, and have narrow $E_p$ distributions. 
However the simulated distributions are still about one order of magnitude 
broader than the $E_p$ distribution observed in the bright BATSE bursts. 
5. Two sub-classes of the innermost model, (h) and (i), have very narrow
$E_p$ distribution (due to much smaller powers of the parameters involved, 
eq.[\ref{Ep6}]). These are comparable to that of the bright BATSE bursts.
However,  it is likely that these emission components contribute only 
partially, and under certain circumstances, to the GRB prompt emission 

\begin{figure}
\centerline{\psfig{file=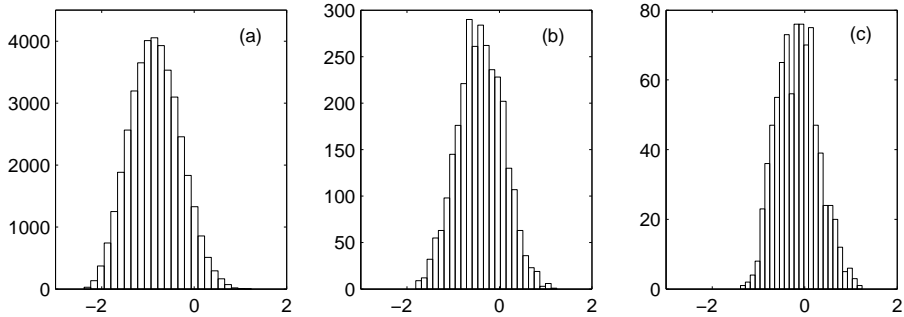,width=12.0cm}}
\caption{The flux-threshold effect on the $E_p$ distribution in the
simulations for Model (d) in Fig.1. (a) all 40000 bursts 
simulated; (b) the 2962 bursts whose fluxes are within the
top three orders of magnitude; (c) the 848 bursts whose fluxes are
within the top two orders of magnitude.}
\label{fig2}
\end{figure}

Several caveats about our simulations are pertinent. First, our primary 
goal here is not trying to match the Preece et al results closely. 
To make such close comparisons, other effects such as the detector
energy band and the instrument response function ought to be taken 
into account, and a much larger parameter space would need to be explored 
for every model. It is also possible that the intrinsic $E_p$ distribution
is broader than what is found by Preece et al. (2000), as indicated by
the growing population of the XRF sources.
Instead, we have adopted a uniform set of parameters for different models, 
which allows an unbiased evaluation among the models. In fact, it may be 
possible to reproduce Preece et al's results by adjusting the distribution 
function and the dispersion of the input parameters within a certain model. 
For example, B\"ottcher \& Dermer (2000) have shown that Preece et al's 
narrow $E_p$ distribution is reproducible even in the $E_p \propto \Gamma^4$
external shock model, by taking into account the instrumental effects
and the flux threshold effect, although they had to fix some input
parameters (e.g. $n_{ext}$). Our Fig.\ref{fig1} nonetheless clearly
reveals the ``relative'' abilities of different models to generate
a narrow $E_p$ distribution. Second, in our simulations most of the
parameters are regarded as independent of each other. However, there
may exist some intrinsic correlations among some parameters, and the
conspiracy among these parameters can reduce the $E_p$ scatter to 
achieve narrower distributions (we have shown this effect in Fig.1a and
1b). More detailed work is needed to explore these intrinsic correlations. 
Finally, it is possible that the flux threshold influences the 
scatter of the distribution. In Fig.\ref{fig2}, we have explored such 
flux threshold effects for a particular model (the internal magnetic
model, same conclusion also applies to other models). The three
subplots indicate different flux thresholds: (a) $E_p$ distribution of
all the 40000 bursts; (b) that of the 2962 bursts whose fluxes are
among the top three orders of magnitude; (c) that of the 848 bursts
whose fluxes are among the top two orders of magnitude. We see that
the $E_p$ distribution indeed gets narrower when raising the flux
threshold, but the effect is only mild. This might be because the
intrinsic $E_p$ scatter due to many unknown parameters has a similar
effect in a very wide flux decades.

\subsection{Nature of X-ray flashes}
\label{sec:xrf}

Following the reports of the identification of the XRFs as a type of 
cosmological explosions resembling GRBs (Heise et al. 2001, 2002; Kippen 
et al. 2001; 2002), several speculations on the nature of these objects 
have been proposed. Whether they smoothly join the GRB distribution or
constitute a related but separate class, the $E_p$ distributions
associated with these objects would depend on the possible models,
some of which we discuss below.

\begin{figure}
\centerline{\psfig{file=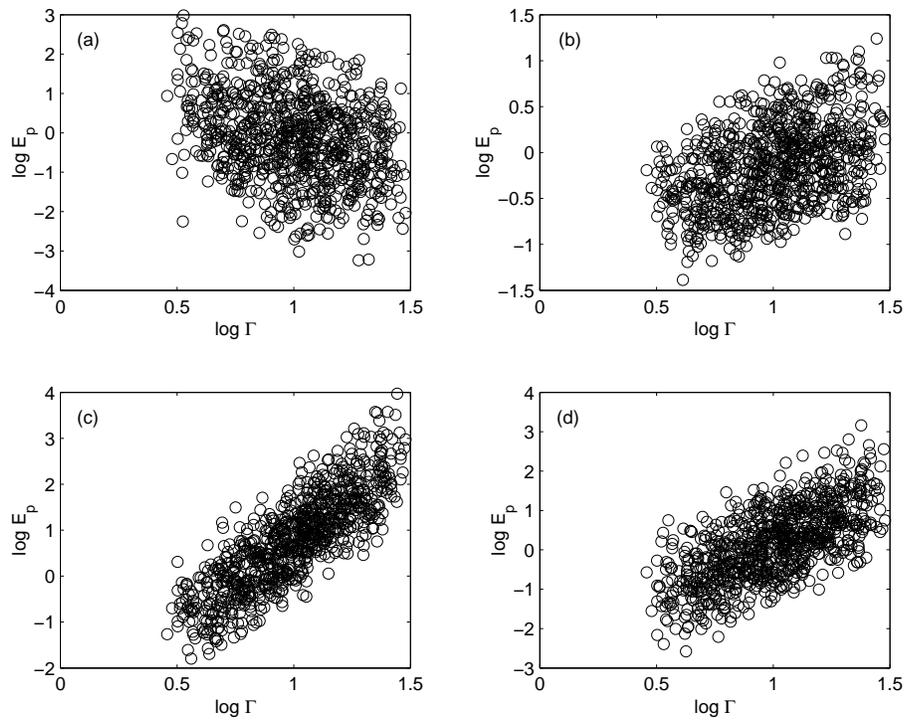,width=12.0cm}}
\caption{Simulated $\log E_p$ as a function of $\log \Gamma$ in
various models: (a) internal shock model; (b) internal magnetic model;
(c) external shock model; (d) external magnetic model. }
\label{fig3}
\end{figure}

1. {\em Dirty fireballs?} Based on the external shock model, Dermer et
al. (1999) discussed a type of then-undiscovered objects with lower bulk
Lorentz factors than conventional GRBs, and suggested that such dirty
fireballs would typically have lower $E_p$ (due to the $E_p\propto
\Gamma^4$ dependence). It is then natural to attribute the XRFs to
such dirty fireballs (Heise et al. 2001). However, we note that a
dirty fireball does not necessarily produce a low $E_p$ burst. In the
optically thin internal shock model, equation (\ref{B}) shows that the 
apparent $\Gamma$ dependence cancels out, and equation (\ref{Ep1}) 
indicates that a dirty fireball has a closer-in internal shock radius, 
hence a higher $E_p$ due to a higher magnetic field (Fig.3a). In the high
$\sigma$ case or the pair-dominated case, and in the external models,
a positive dependence of $E_p$ on $\Gamma$ is retained, and a dirtier
fireball tends to produce a lower $E_p$ (Fig.\ref{fig3}).

2. {\em High redshift bursts?} An interesting possibility is that XRFs
are GRBs at much higher redshifts (e.g. $z>6$), which may thus be
related to the death of the first stars in the universe (Heise et al.
2001). However,  some of the other collateral evidence for high-$z$
location, e.g., time dilation for both the total durations and the
individual pulses, is lacking, which casts some doubts on such an
interpretation (Lloyd-Ronning 2002). This problem is made less severe
by the fact that also in normal GRBs, the presence of such time dilation 
is less noticeable than the intrinsic dispersion, and becomes noticeable 
only when analyzing very large samples (e.g. Norris et al. 2000).

3. {\em Off-beam GRBs?} There is now strong evidence that at least a
fair fraction of GRBs are collimated. This raises the possibility of
interpreting the XRF phenomenon as being due to viewing angle effects.
In a version of such models where a sharp jet edge is assumed (Yamazaki,
Ioka \& Nakamura 2002), XRFs are those off-beam GRBs in which the observer
misses the bright jet cone. Such a model requires the XRFs to be very
nearby, $z \siml 0.2$. The redshift measurement of one X-ray rich GRB
011211 (a close relative of XRFs) at $z=2.14$ seems to suggest that at
least some XRFs may not be interpreted in this scenario. In the
universal jet beam model (Rossi et al. 2002; Zhang \& \Mesz~2002), the
off-beam population is greatly reduced, but bursts observed at large
viewing angles tend to be ``dirty''. The possible connection of such a
configuration to XRFs (within the framework of the collapsar models)
has been suggested by Woosley et al. (2002 and references therein).

4. {\em Photosphere-dominated fireballs?} \Mesz~ et al. (2002) suggest
that XRFs may be accounted for within the standard fireball internal
shock model with moderate redshifts, due to a dominant contribution of
either the baryonic or the shock pair photosphere. A similar proposal in
the high-$\sigma$ case has also been suggested (Drenkhahn 2002).

The first two speculations invoke the dispersion of one particular
parameter, while the total $E_p$ dispersion is the combined
dispersions of many independent parameters. Whether the influence of
one particular parameter is prominent depends on how many independent
parameters are involved, and what is the power of that particular
parameter in the $E_p$ expression. To show this effect, we plot in
Fig.\ref{fig3} the predicted $E_p$ as a function of $\Gamma$ in four
models: (a) internal shock model; (b) internal magnetic model; (c)
external shock model; and (d) external magnetic model. 
We find that the $E_p$ broadening due to the $\Gamma$ scatter is more 
evident for the external models in which high powers of $\Gamma$ are 
invoked. Such an effect also depends on the bandwidth of the detector. 
Only when the bandwidth is larger than the intrinsic scatter at a same 
$\Gamma$ could the broadening due to $\Gamma$ scatter be noticeable. 
For the same reason, the redshift scatter only contributes weakly to
the $E_p$ scatter, so that the high redshift interpretation of XRFs 
is not quite plausible. The photosphere interpretation in fact invokes
possibly two (or even more) emission components with different $E_p$
distributions. A natural expectation of this model is that the $E_p$
distribution of the GRB/XRF combined population may have a
non-lognormal or even bimodal distribution, which seems not incompatible 
with the current small sample (Kippen et al. 2002). To identify the 
nature of the XRFs, clearly more data are needed, including both 
redshift measurements and spectral fits.

\subsection{$E_p-L$ correlation}
\label{sec:epkLcorr}

The positive correlation $E_p \propto L^\delta$, (Amati et al. 2002;
Lloyd-Ronning \& Ramirez-Ruiz 2002) also poses interesting constraints 
on the models. It is seen in Table \ref{tbl-2} that the different models 
give different predictions for the $E_p-L$ dependence.  The complications
arise from the $E_p$ dependence on the unknown $\Gamma$ expected in 
these various models, since there is no simple way to relate $\Gamma$ to 
the observables.
In the above Monte Carlo simulations, we have adopted a positive
$\Gamma \propto L^k$ ($k=1/2$ specifically) correlation, which is
consistent with the expectation of the theoretical models (Kobayashi
et al. 2002; Rossi et al. 2002; Zhang \& \Mesz~2002; Salmonson \&
Galama 2002).
If one takes $k$ as a free parameter, some constraints on $k$ may be 
imposed by the observed $\delta$.  For example, the internal shock 
synchrotron model could not give a positive $E_p-L$ correlation if 
$k > 1/4$. Ramirez-Ruiz \& Lloyd-Ronning (2002) also noticed this, 
and argued that their observed relation (Lloyd-Ronning \& Ramirez-Ruiz 
2002) can be only accommodated within the IC-dominated model with an 
additional assumption of a $\gamma_e \propto \Gamma$ (the $\Gamma^{-2}$
dependence thereby canceling out). However, we note that the
IC-dominant picture is less favored for other reasons as discussed
above. Similar analyses could be made for the other models. For
example, both external models seem to predict a too steep $E_p-L$
correlation, unless the $\Gamma-L$ relation is rather mild. The
internal magnetic model or the pair dominated model are however
compatible with the data for reasonable parameters.

\section{Conclusions}
\label{sec:conc}

The nature of the GRB prompt $\gamma$-ray emission, including the
emission site and the energy contents of the fireball, are still poorly
known after more than 30 years of efforts. We have analyzed the various
fireball model variants within a unified picture, and have revisited
the $E_p$ predictions of different models. These models are tested
against the current GRB spectral data with a simple Monte Carlo
simulation, with attention to the narrowness of the distribution and
its dependence on some of the physical parameters. Our aim is to set up 
a general theoretical framework to allow unbiased tests of these models
against the known data. Based on the analysis in this paper, we
can evaluate the existing fireball models as follows.

1. The internal shock model is generally regarded as the most
attractive candidate for the prompt $\gamma$-ray emission of
classical GRBs. The highly variable, spiky GRB lightcurves are
naturally reproduced in such a model, and many studies have shown 
that it is successful in reproducing many of the GRB properties 
(e.g. Kobayashi, Piran \& Sari 1997; Daigne \& Mochkovitch 1998;
Panaitescu, Spada, \Mesz~ 1999; Spada, Panaitescu, \Mesz~2000;
Ramirez-Ruiz \& Fenimore 2000; Guetta et al. 2001). 
Our findings in this paper indicate two important caveats for the 
internal shock model. First, unless the dispersions in the shock 
parameters (e.g. $\epsilon_e$, $\epsilon_B$, $\theta_p$, $p$, etc) are 
very small or there exist some intrinsic correlations among the 
parameters, the internal shock model generates an $E_p$ distribution 
(Fig.1a) which, even in the optimistic case, is at least one order of 
magnitude broader than the straight (i.e. not otherwise corrected)
distribution given by Preece et al. (2000). The calculated distribution
may be still compatible with the data if XRFs are included in the
observations, which intrinsically broaden the $E_p$
distribution\footnote{It is worth mentioning that within the same
burst, the internal  
shock model also tends to produce a wide $E_p$ distribution, which is 
incompatible with Preece et al. (2000) unless some fine tuning is made 
(S. Kobayashi, 2002, in preparation).}. 
Second, the synchrotron internal shock model may not be able to 
reproduce the $E_p-L$ positive dependence unless the $\Gamma$ 
distribution is un-related or weakly related to $L$, or some further 
assumption is made (e.g. Ramirez-Ruiz \& Lloyd-Ronning 2002).

2. The high-$\sigma$ internal model could inherit most of the merits
of the internal shock model, with some additional advantages such as a 
smaller $E_p$ dispersion and the right $E_p-L$ correlation. The caveat in
such models is that they have so far been less well-studied than the 
internal shock models, including the basic particle acceleration and 
emission processes. It is also unclear how to lower the typical $E_p$ 
energy to the sub-MeV band, and the physical context of how a high-$\sigma$
flow is launched in a collapsar is not well explored. More investigations 
in this direction are desirable (e.g. Blandford 2002).

3. In both the shock and magnetic dissipation internal models, both
a narrow $E_p$ distribution and the right $E_p-L$ correlation are
attainable if a pair photosphere is formed. However, it is unclear how
common such a situation would be. The sharp lightcurve spikes may be also
hard to reproduce.

4. The prompt $\gamma$-ray emission predicted in external models is
expected in most cases, unless the external medium is very under-dense
or previous internal dissipation of the bulk kinetic energy has been highly
efficient. Whether the observed GRB prompt emission is attributable to
this component is in question. Our simulations show that these models
produce too broad an $E_p$ distribution compared with other models,
and probably a too steep $E_p-L$ correlation. Other arguments against
such scenarios include the need for additional assumptions (blobs) and
the inefficiency involved in interpreting the variability (Sari \&
Piran 1997, but see Dermer \& Mitman 1999). An important feature of
these models is that $E_p$ are positively dependent on the ambient ISM
density $n_{ext}$, which is in principle a measurable parameter. This
provides a potential test for the model. More broadband afterglow fits
are needed before a significant statistical evaluation can be made. 
Even if the external scenario does not account for the GRB prompt emission, 
studies of such a model are nevertheless meaningful since they can explore
the important bridge between the prompt emission and the afterglow.

5. The emission component coming directly from the baryon photosphere
is expected to partially contribute to the observed GRB prompt emission. 
The $E_p$ distributions for different regimes are narrow, and $E_p-L$ 
correlations are easily accommodated at least in the regime where the 
emission is the strongest (i.e., $\eta>\eta_{c2}$). The best guess is 
that such a component appears mixed in with other components, and becomes 
important under certain conditions.

6. All the IC models tend to generate broader $E_p$ distributions as
compared with their synchrotron counterparts. They are also less
favored due to other reasons.

Finally, our entire discussion in this paper is within the context 
of a naked central engine. A more realistic scenario invokes the
fireball-progenitor envelope interaction, e.g. in collapsar models, 
which is receiving increased attention. These would lead to additional
emission components, which are beyond the scope of the current paper.

\acknowledgments{We thank S. Kobayashi, N. M. Lloyd-Ronning, and
E. Ramirez-Ruiz for valuable discussions and comments, and the referee
for a detailed report that helped to improve the presentation of the
paper. We also acknowledge useful conversations or correspondence with
R. D. Blandford, Z. G. Dai, C. D. Dermer, J. Heise, D. Kazanas,
M. Kippen, and A. I. MacFadyen. This work is supported by NASA
NAG5-9192 and NAG5-9153.}

\newpage
\begin{deluxetable}{ll}
\tablecaption{Fireball variants \label{tbl-1}}
\tablewidth{0pt}
\tablehead{
\colhead{Condition} & \colhead{Fireball properties}}
\startdata
$\sigma \siml \sigma_{c1}$ & kinetic energy dominated, strong shocks
 possible \\
$\sigma_{c1} < \sigma \leq \sigma_{c2}$ & Poynting dominated, MHD
 does not break globally, but may break locally, no strong shocks \\
$\sigma_{c2} < \sigma \leq \sigma_{c3}$ & completely Poynting dominated,
 MHD breaks globally, no strong shocks, baryon electrons non-negligible  \\
$\sigma > \sigma_{c3}$ & completely Poynting dominated, MHD breaks
 globally, no strong shocks, baryon electrons negligible \\
\hline
\enddata


\end{deluxetable}

\begin{deluxetable}{llll}
\tablecaption{Model predictions for $E_p$ \label{tbl-2}}
\tablewidth{0pt}
\tablehead{
\colhead{Models} & \colhead{Sub-categories}   & \colhead{$E_p$}
 & \colhead{Comments}}
\startdata
 & low-$\sigma$: internal shocks
 & $\propto \epsilon_{x3} L^{1/2} \Gamma^{-2} t_v^{-1} (1+z)^{-1}$
 & $\gamma_e^{(1)}$ relevant \\
Internal models & high-$\sigma$: magnetic dissipation & $\propto
 \epsilon_{x5} \Gamma (1+z)^{-1}$ & $\gamma_e^{(2)}$ relevant \\
 & pair photosphere & $\propto \Gamma T'_\pm/{\tau'}_\pm^2 (1+z)^{-1}$
 & Comptonized spectrum \\
\hline
External models & low-$\sigma$: external shocks
 & $\propto \epsilon_{x6} \Gamma^4 n_{ext}^{1/2} (1+z)^{-1}$
 & $B$ generated in-situ \\
& high-$\sigma$: plasma-barrier interaction &
 $\propto \epsilon_{x7} \Gamma^{8/3} L^{1/2} E^{-1/3}
 n_{ext}^{1/3}(1+z)^{-1}$  & $B$ carried from the wind \\
\hline
& $\eta<\eta_{c1}\sim 250(1+\sigma)^{-1/5}$ & $\propto
 L^{-5/12}t_{v,m}^{1/6}\Gamma^{8/3}(1+z)^{-1}$ & wind coasting regime \\
Innermost models & $\eta_{c1}< \eta<\eta_{c2}$ & $\propto L^{-1/12}
 t_{v,m}^{-1/6}\Gamma (1+z)^{-1}$ &  shell coasting regime \\
& $\eta>\eta_{c2}\sim 10^4(1+\sigma)^{-1/3}$ & $ \propto
 L^{1/4}t_{v,m}^{-1/2}(1+z)^{-1}$ & shell acceleration regime \\
\hline
\enddata


\end{deluxetable}

\newpage

\end{document}